\documentclass[onecolumn,10pt]{article}
\usepackage[top=.75in, bottom=.75in, left=.75in, right=.75in]{geometry}
\usepackage{amsmath,amsfonts,amscd,amssymb}
\usepackage{graphicx}
\usepackage{epstopdf}
\usepackage{overpic}
\usepackage{cancel}
\usepackage{rotating}
\usepackage{url}
\usepackage{caption}
\usepackage{color}
\usepackage{rotating}
\usepackage{multirow}
\usepackage{wrapfig}
\usepackage{mathtools}
\usepackage{subeqnarray}
\usepackage{setspace}
\usepackage{palatino} 
\usepackage[numbers,sort&compress]{natbib}

\usepackage[bottom,flushmargin,hang,multiple]{footmisc}
\usepackage{lipsum}
\newcommand\blfootnote[1]{%
  \begingroup
  \renewcommand\thefootnote{}\footnote{#1}%
  \addtocounter{footnote}{-1}%
  \endgroup
}

\definecolor{header1}{cmyk}{0,0,0,1}

\usepackage[utf8]{inputenc}

\usepackage[normalem]{ulem}
\usepackage{color}

\title{\vspace{-.45in}{\huge\selectfont \textbf{Emerging trends in machine learning for computational fluid dynamics}}\vspace{-.15in}}

\author{\normalsize{Ricardo Vinuesa$^{1,2*}$ and Steven L. Brunton$^{3}$}\\
\footnotesize{$^1$ FLOW, Engineering Mechanics, KTH Royal Institute of Technology, Stockholm, Sweden}\\
\footnotesize{$^2$ Swedish e-Science Research Centre (SeRC), Stockholm, Sweden}\\
\footnotesize{$^3$ Department of Mechanical Engineering, University of Washington, Seattle, WA 98195, United States \vspace{-.2in}}
}

\date{}

\begin{document}
\maketitle

\blfootnote{$^*$ Corresponding author: Ricardo Vinuesa (rvinuesa@mech.kth.se)}
\vspace{-.2in}
\begin{abstract}

The renewed interest from the scientific community in machine learning (ML) is opening many new areas of research. 
Here we focus on trends in ML that are providing opportunities to advance the field of computational fluid dynamics (CFD). We discuss synergies between ML and CFD that have already shown benefits, and we also assess areas that are under development and may produce important benefits in the coming years. 
We believe that it is also important to emphasize a balanced perspective of cautious optimism for these emerging approaches. \\
    

\end{abstract}

\section*{Introduction}\label{sec:intro}

Machine learning (ML) is a rapidly developing field of research that has transformed the state-of-the-art capabilities for many traditional tasks in computer science, such as image classification and captioning, natural language processing, and recommender systems.   
The numerous success stories of ML have led to widespread adoption in the scientific and engineering communities as well, fueled by a growing wealth of data, computing resources, and advanced optimization algorithms. This is especially true in the field of fluid mechanics, where emerging technologies complement existing computational and experimental methods, providing a unified approach to building models from data~\cite{Brunton2020arfm}.    

Despite these advancements, there remains a gap in understanding how ML can be best integrated with computational fluid dynamics (CFD). This paper aims to explore the synergies between ML and CFD, showcasing the potential benefits and challenges in combining these fields. ML can advance CFD in areas such as turbulence modeling, development of inflow boundary conditions, subgrid-scale models for large-eddy simulations (LES), closures for Reynolds-averaged Navier--Stokes (RANS) equations, development of reduced-order models (ROMs), and flow control~\cite{vinuesa_brunton}. Our approach is to first examine established techniques, such as proper-orthogonal decomposition (POD) and dynamic-mode decomposition (DMD), alongside deep-learning techniques with autoencoders. Next, we delve into emerging opportunities where ML and CFD can be further integrated, highlighting ongoing challenges and potential solutions. We conclude by summarizing the insights gained and potential future directions for this interdisciplinary research.

\section*{Overview of applications}

Some of the more relevant applications at the intersection of ML and CFD include the following:

\begin{itemize}
\item {\bf Modeling the near-wall region of wall-bounded turbulence.} Milano~and~Koumoutsakos~\cite{milano_koumoutsakos} developed, 20 years ago, a method to predict the relevant features of turbulent channel flow close to the wall by using deep neural networks. In addition to the interest of this work from the perspective of modeling turbulence, and potentially developing wall models, the authors also established connections between neural networks and traditional methods such as proper-orthogonal decomposition (POD). 
In particular, they showed that by restricting the neural network model to be linear, the network essentially learns the features produced by standard POD. 
This topic has received renewed attention in recent years due to the possibility of training deeper neural network models with larger data sets, as well as the emergence of novel learning architectures~\cite{bae_koumoutsakos}. 

\item {\bf Development of inflow conditions for turbulence simulations.} This is a critical area for CFD when it comes to achieving high-Reynolds-number ($Re$) conditions and simulating complex geometries. Spatially-developing turbulent boundary layers (TBLs) require very long domains to reach high Reynolds numbers, and if part of the low-$Re$ region can be replaced by an adequate inflow condition, simulations at the relevant high-$Re$ regime can become feasible. Similarly, if a simulation is designed to study the turbulent flow around a complex array of obstacles, it is beneficial to replace the inflow section of the simulation by a suitable inflow condition, thus yielding significant computational savings. Traditional approaches to these inflows have relied on various filters, recycling approaches, and imposing synthetic turbulent structures. However, recent progress in ML has enabled the use of deep-learning approaches to produce robust inflow conditions, including the use of modern architectures, such as transformers~\cite{heechang}. 

\item {\bf Development of boundary conditions for external-flow simulations.} 
One of the challenges of simulating external flows, {\it e.g.} the flow past an airfoil, is the need for large computational domains to adequately reproduce the far-field conditions and the pressure distribution around the wing. While the resolution of the portion of the domain near the far field is often already quite coarse, being able to replace this region entirely by a suitable boundary condition would lead to significant computational savings. Furthermore, it is important to be able to establish prescribed pressure-gradient (PG) conditions so as to study PG TBLs and their non-trivial flow-history effects. To this end, a number of ML approaches have been recently developed for effective boundary-condition design~\cite{bae_koumoutsakos}. In particular, Gaussian-process regression has been successfully employed to optimize the shape of the domain to prescribe a particular PG distribution. 

\item {\bf Improved subgrid-scale (SGS) models for large-eddy simulations (LESs).} 
In this application, the goal is to develop frameworks to supplement coarse numerical simulations, where only the largest turbulent scales are resolved, with additional information corresponding to the flow scales that are too fine to be properly simulated with the chosen computational mesh~\cite{duraisamy_et_al}. This approach leads to significant computational savings, but opens questions regarding the physical relevance and accuracy of the assumptions from the SGS model. 
One way in which ML is helping to develop SGS models is to use filtered direct-numerical-simulation (DNS) data to obtain the information required to supplement the coarse model, and then to train data-driven algorithms to predict this filtered field. 
However, there are limitations to this approach, and recently, more nuanced approaches have been developed that replace supervised learning with reinforcement learning. This also includes innovative approaches to develop wall models~\cite{bae_koumoutsakos}, which can enable high-$Re$, wall-bounded turbulence studies, as discussed in more detail below.

\item {\bf Enhanced closures for the Reynolds-averaged Navier--Stokes (RANS) equations.} 
Beyond traditional approaches, based on using high-fidelity data to fit the parameters from the eddy-viscosity-based Reynolds-stress models, ML has brought a number of additional approaches to RANS modeling~\cite{duraisamy_et_al}. 
These approaches include the discovery of novel closure functional forms and interpretability of data-driven turbulence models. 
Furthermore, recent approaches based on physics-informed neural networks (PINNs)~\cite{Raissi2019jcp} have also led to successful predictions of the Reynolds stresses in the context of RANS. We elaborate on these topics below.

\item {\bf Robust methods for non-intrusive sensing and super-resolution.} 
To study and control turbulent flows experimentally, non-intrusive sensing is important, for example being able to predict the flow based on measurements at the wall. 
This is an area where numerical simulations help in the design of better experiments, and where data-driven methods have traditionally played an important role. 
Typical methods to predict the flow based on wall information include linear stochastic estimation (LSE), extended proper-orthogonal decomposition (EPOD), and system-identification approaches based on transfer functions. 
The main limitation of these methods is that they are linear, and although they are capable of accurately predicting the linear-superposition mechanism of turbulence, they do not capture nonlinear modulation. 
Deep-learning methods, however, are capable of accurately reproducing the nonlinear scale interactions in turbulent flows, and have recently exhibited superior performance compared with linear methods. 
In particular, convolutional neural networks (CNNs), generative adversarial networks (GANs), and other computer-vision architectures provide new techniques to perform non-intrusive sensing in turbulence, even with coarsely-sampled wall information~\cite{fukami2019super}.
\item {\bf Novel approaches for flow control.} Finally, another area where ML is starting to have a significant impact is flow control.  
Again, numerical simulations have often been used to obtain robust control strategies that can subsequently be deployed in experimental studies. 
Active flow control, in which the actuation requires energy input, has been extensively used in the context of turbulent flows, both to reduce drag and to enhance mixing, depending on the application. 
Closed-loop control is typically the most effective approach, because it uses measurements of the instantaneous state of the flow to dynamically determine the control action; note the importance of sensing mentioned above. 
One traditional method of closed-loop control is opposition control, in which a synthetic jet at the wall actuates with the opposite sign of the wall-normal fluctuations in the near-wall region. 
ML methods are enabling more sophisticated control strategies, {\it e.g.} through genetic programming (with the advantage of providing interpretable control policies), or via deep reinforcement learning, which is enabling the discovery of novel control approaches in a wide range of flows~\cite{rabault2019artificial}, including turbulence control~\cite{guastoni2023deep}.
\end{itemize}

In the remainder of this contribution, we will focus on established techniques and emerging trends, where new applications of ML algorithms have the potential to significantly impact numerical simulations of fluid flows. After assessing the synergies between ML and CFD, we will summarize our conclusions and provide an outlook.

\section*{Synergies between machine learning and computational fluid dynamics}\label{sec:body}

The intersection of machine learning and computational fluid dynamics is a rapidly growing discipline.  
Therefore, we have organized the following into established methods and connections, emerging and developing research areas, and ongoing challenges and limitations. 

\subsection*{Established techniques}

New approaches in ML are having a significant impact in the development of ROMs. 
Even the most complex flows exhibit certain patterns and dominant structures, which can be used to construct models with simpler dynamics than the original physical system, but containing some of its essential ingredients~\cite{Benner2015siamreview,Rowley2017arfm,Brunton2020arfm}. 
These patterns enable the development of ROMs, which can be used to make certain predictions of the complete system with lower fidelity, at a significantly lower computational cost. Some of the most widely-used methods for ROM development rely on linear algebra. For instance, the well-known proper-orthogonal decomposition (POD), which is also known as principal-component analysis (PCA), is based on the singular-value decomposition (SVD) algorithm for matrices. 
POD reduces the dimensionality of flow-field data by identifying the most important modes of variation in the data, which are typically associated with large-scale features in the flow. We assume that a velocity field $\pmb{u}(\pmb{x},t)$ can be expressed as a linear combination of the mean flow $\overline{\pmb{u}}(\pmb{x})$ and a set of orthogonal spatial modes $\pmb{\phi}_i(\pmb{x})$:
\begin{equation}\label{eq:pod}
\pmb{u}(\pmb{x},t)=\overline{\pmb{u}}(\pmb{x})+\sum_{i=1}^N a_i(t) \pmb{\phi}_i(\pmb{x}),
\end{equation} 
where $a_i(t)$ are the temporal coefficients specifying the amplitudes of the spatial modes, $\pmb{x}$ are the spatial coordinates, $t$ is the time, and boldface denotes vectors. To find the temporal coefficients $a_i(t)$ in equation~(\ref{eq:pod}), we can use the snapshots of the flow-field data $\pmb{u}(\pmb{x}_j,t_k)$ at discrete spatial locations $\pmb{x}_j$ and time instances $t_k$. These snapshots are organized into a matrix $\pmb{U}=\left [ \pmb{u}_{jk} \right ]$, where each column corresponds to a time instant and each row corresponds to a spatial location. We then perform singular-value decomposition (SVD) on the matrix $\pmb{U}$, obtaining a set of orthonormal spatial modes $\pmb{\phi}_i(\pmb{x})$ and their corresponding singular values $\sigma_i$, related to the energy contained in each mode. We keep $N$ modes by choosing a threshold for the singular values that captures a desired percentage of the total energy in the system.

Substituting the POD into the governing equations of the fluid flow, such as the Navier--Stokes equations, gives a system of equations for the coefficients. This is known as Galerkin projection, and it involves projecting the governing equations onto the subspace spanned by the POD modes. The resulting is a system of ordinary differential equations for the temporal coefficients $a_i(t)$, a dynamical system representing the dynamics of the flow restricted to the low-dimensional POD subspace.

Two important properties of the POD modes are their orthogonality, implying that a particular physical feature of the system will be represented in only one mode of the expansion, and the fact that they are ranked by their energetic contribution to $\pmb{u}(\pmb{x},t)$, thus facilitating the truncation process. 
Note that equation~(\ref{eq:pod}) decomposes the spatio-temporal problem into spatial modes and their temporal dynamics, so that POD is fundamentally based on a separation of variables.  
Thus, having $\pmb{\phi}_i(\pmb{x})$ implies that only the temporal coefficients need to be obtained in order to build a ROM.  

A number of ML-based approaches have been used to predict the temporal coefficients in a purely data-driven manner, including the various types of recurrent neural networks~\cite{srinivasan2019predictions}, Koopman-based frameworks~\cite{eivazi2020recurrent}, sparse identification of nonlinear dynamics~\cite{Brunton2016pnas}, convolutional neural networks~\cite{sole_conv} and even more recently transformers~\cite{heechang}. These data-driven approaches have enabled the development of robust ROMs for flow cases of significant complexity. Another data-driven approach to identify flow features is the dynamic-mode decomposition (DMD)~\cite{Rowley2009jfm,Schmid2010jfm,dmd_tu}, in which the spatio-temporal data can be decomposed into Fourier-like modes with their corresponding amplitudes, frequencies, and growth rates. DMD differs from POD in several ways. While POD decomposes the data into \emph{spatial} modes and temporal coefficients, DMD decomposes it into \emph{dynamic} oscillatory modes with a single frequency. DMD is a purely data-driven approach, and does not require an equation as is needed in Galerkin projection. DMD captures transient dynamics, while POD is more suited for quasi-steady flow. Both are powerful tools for reduced-order modeling, and the choice between them depends on the problem at hand.

The methods discussed above learn a linear subspace where the dynamics of the ROM are expressed. Deep learning enables learning a nonlinear coordinate system on a curved manifold, where it is possible to compactly represent the complex nonlinear dynamics of fluid-flow systems~\cite{Floryan_Graham}. In a neural network, each layer produces an output by taking a linear combination of the inputs, adding a bias, and passing it through a nonlinear activation function. When we consider a linear neural network ({\it i.e.}, the activation function is linear) with an identity mapping, we find that the matrix of weights is equivalent to the matrix formed with the eigenfunctions of POD, implying that the linear operations used in POD can be seen as the operations of a linear neural network. We can obtain the POD via training a linear neural network using backpropagation with the squared distance between the original and reconstructed data as the cost function. As discussed above, Milano and Koumoutsakos~\cite{milano_koumoutsakos} used this idea to develop a nonlinear generalization of POD by adding a nonlinear activation function in two hidden layers of the neural network. They used this model for reconstruction of the near-wall flow in a turbulent channel, finding it gives better compression than linear POD (better reconstruction for unseen data), at a higher computational cost. In the nonlinear generalization of POD using a neural network, one is replacing SVD by backpropagation to learn a lower-dimensional representation of the data. This low-dimensional representation can be considered as the nonlinear principal components of the data, capturing the essential features of the flow. The neural-network model can then be used to reconstruct the high-dimensional fluid-flow data from the low-dimensional representation. In deep learning, this two-step process is called an autoencoder: a network that projects high-dimensional data into and from a lower-dimensional latent space. In an autoencoder, the so-called encoder $\mathcal{E}$ maps the input data $\pmb{u}$ to a low-dimensional latent space $\pmb{r}$, {\it i.e.} $\pmb{u} \rightarrow \pmb{r}$, while the decoder $\mathcal{D}$ maps the latent space back to the original data: $\pmb{r} \rightarrow  \tilde{\pmb{u}}$, obtaining the reconstruction $\tilde{\pmb{u}}$. When the autoencoder produces an accurate reconstruction, it also provides an effective low-dimensional representation of the original data in the latent space. The process for implementing an autoencoder is summarized below:
\begin{subequations}
\begin{equation}\label{eq:ae}
\mathcal{F}=\mathcal{D} \circ \mathcal{E}, 
\end{equation}
\begin{equation}\label{eq:ae2}
\tilde{\pmb{u}}=\mathcal{F} (\pmb{u};\pmb{w}), 
\end{equation}
\begin{equation}\label{eq:ae3}
\mathcal{L}_{\rm rec}=\varepsilon (\pmb{u},\tilde{\pmb{u}}). 
\end{equation}
\end{subequations}
Here $\mathcal{F}$ denotes the network, $\pmb{w}$ its parameters to be optimized through training, $\varepsilon$ is the loss function, and $\mathcal{L}_{\rm rec}$ is the reconstruction loss. 
Through this data-driven framework, it has been possible to obtain nonlinear modal decompositions of a number of flows, leading to very compact representations of the original data in the latent space due to the inherent nonlinearity of the method.

Autoencoders, unlike POD, do not inherently possess the two important properties of orthogonality and rank ordering by the energy content. Recent work has addressed both issues. Obtaining the nonlinear modes in ranked order of their energy content is possible with hierarchical encoders, which leverage convolutional layers to learn increasingly abstract features~\cite{fukami_autoencoders}. They thus can exploit spatial locality and multi-scale information in the flow input data. The hierarchical formulation starts by using a network, denoted as $\mathcal{F}_1$, to map the original data into a one-dimensional latent vector: $\pmb{r}_1=\mathcal{E}_1(\pmb{u})$. Next, a second network $\mathcal{F}_2$ is trained to map the input data $\pmb{u}$ into a two-dimensional latent space, defined as: $[\pmb{r}_1,\pmb{r}_2]$, where $\pmb{r}_1$ is already known and $\pmb{r}_2$ is obtained at this level. This recurrent relationship of networks can be extended to higher dimensions, building a ROM where each additional latent vector contributes less to the reconstruction of the original field. To promote orthogonality, disentanglement among the latent vectors can be favored by adding a $\beta$-weighted penalty term to the loss function in varitational autoencoders ($\beta$VAEs). In tests using a complex turbulent flow~\cite{ae_modal}, this methodology provided around $90\%$ energy reconstruction with just 5 autoencoder modes, whereas 5 POD modes would yield a reconstruction of about $30\%$ of the energy. These results highlight the potential of using autoencoders to develop very compact ROMs, even in turbulent flows. 

Established ML techniques are playing an important role in developing improved RANS models~\cite{duraisamy_et_al}. While traditional approaches focus on fitting coefficients in classical RANS strategies, researchers are investigating innovative methods interesting new approaches for model discovery. These methods include genetic programming~\cite{cranmer_et_al} or sparse identification of nonlinear dynamics (SINDy)~\cite{Brunton2016pnas}. The key advantage of these approaches is their ability to provide interpretable models that are not constrained by the functional forms of existing ones. Interpretability is particularly important, as many deep-learning-based approaches that demonstrate potential for novel RANS models are not interpretable. Cranmer {\it et al.}~\cite{cranmer_et_al} present a promising approach to incorporating interpretability into RANS models developed through deep learning. In their work symbolic equations are derived through symbolic regression (using genetic programming) based on the input-output behavior of a trained neural-network model. These models demonstrate performance equivalent to the original neural network, while offering enhanced interpretability and generalizability properties. 

Machine learning is showing great potential in other modelling tasks such as assessing the impact of compressibility on the flow~\cite{xiao_comp}, and evaluating the properties of stratified flows~\cite{9433573}, even at large scales. Measurement techniques are also being improved through the integration of machine learning~\cite{ai4exp}, particularly when combined with simulations.

\subsection*{Emerging opportunities}

Large-scale simulations of turbulent flows, particularly at high $Re$, are computationally expensive due to the multi-scale nature of turbulence. With increasing $Re$, where the separation of scales increases and the smallest scales near the wall become progressively smaller. Direct numerical simulation (DNS) requires computational meshes fine enough to resolve all relevant flow scales, leading to a growing number of required grid points proportional to $Re^{37/14}$ (or $Re^{13/7}$ for large-eddy simulation, LES)~\cite{choi_moin}. As a result, simulating industrially-relevant Reynolds numbers remains infeasible with current high-performance-computing (HPC) facilities. 
Reducing the computational cost of these simulations is critically important for the field of CFD, particularly for turbulent flows in conditions close to full-scale applications. Kochkov~{\it et al.}~\cite{hoyer} proposed an interesting approach for accelerating simulations using ML. Their method involves reducing the resolution of the numerical mesh, which automatically lowers computational cost. 
However, using a coarse mesh with conventional numerical methods often gives incorrect physics and erroneous results. 
Kochkov~{\it et al.}~\cite{hoyer} propose a deep-learning framework to establish a correction between the low-cost coarse-resolution simulation and the costly fine-resolution simulation. 
Their results show that, even with computational meshes 8--10 times coarser in each spatial direction, it is still possible to recover the most relevant features of the flow, including some much smaller than the grid spacing in the coarse case, yielding excellent agreement with the full-resolution reference data. 
Their strategy involves filtering the fine-resolution data to produce a coarse-resolution dataset for the same time steps as the original mesh. A convolutional neural network (CNN) is then trained to predict the residual between the coarse and fine simulations at each step, effectively learning how to supplement the coarse simulation with missing information. CNNs are widely used in computer vision because they can exploit spatial correlations in the data. The vortical flow used in this study exhibits coherent structures that can be effectively predicted using the various convolutional filters in the CNN. Early hidden layers focus on simpler and more abstract features, while layers closer to the output can reproduce and synthesize more complex features due to the concatenated application of convolutions of this architecture. 
This hierarchical representation of structures is well suited to the multi-scale nature of turbulent flows, making CNNs a natural choice for such predictive tasks. In general, super-resolution techniques have shown great promise in highly-structured, multi-scale fluids applications~\cite{fukami2019super}.  

It is important to note that the flow case used in Kochkov et al.~\cite{hoyer} is relatively simple: the two-dimensional Kolmogorov flow sustained by an external-forcing term. 
While the results in this flow case are quite promising, it is crucial to assess the feasibility of this approach in more complex scenarios, such as three-dimensional, unforced turbulence, where fluctuations are maintained through properly resolving the near-wall region, where most of the production occurs up to moderate Reynolds numbers. 
Successfully accelerating such a flow simulation would be a significant milestone in CFD and the study of high-$Re$ turbulence. 
In a related context, the neural-operator framework~\cite{li2020neural} is a promising approach, where a generalization of neural networks is proposed to learn operators capable of approximating complex nonlinear operators. This has already been shown to perform well for coarse meshes in two dimensions, at a reduced computational cost compared to traditional partial differential equation (PDE) solvers.

Using coarser meshes and supplementing the resolved flow with an ML-based term to recover unresolved information can also be combined with a wall model. In this approach a model is used to calculate the wall-shear stress, eliminating the need to resolve the computationally expensive area close to the wall and yielding significant computational savings, particularly at higher Reynolds number. This approach is popular in atmospheric boundary layers (ABLs)~\cite{stoll_et_al}, where extreme flow conditions typically require such an approach for realistic Reynolds numbers. An alternative method involves exploiting the self similarity of turbulent structures within the overlap region for a more sophisticated condition~\cite{mizuno_jimenez}. While promising, current wall models models have limitations and can benefit from ML to devise more advanced strategies. Some studies use CNNs~\cite{ari_etmm} to predict fluctuations at the boundary-condition plane based on information from another plane. Although these methods show better predictions, missing information and scales at the input plane hinder the development of robust models that perform well a posteri. An interesting direction involves using reinforcement learning (RL) to predict the most-likely output plane based on the input plane, leading to adequate flow features and turbulence statistics~\cite{bae_koumoutsakos}. The advantage of this approach is its unsupervised nature, not requiring labelled data for the boundary condition. The most suitable boundary condition may not be the filtered DNS plane; thus, exploring alternatives given a predefined reward function could be a viable approach for these simulations.

\subsection*{Ongoing challenges}

Several aspects of CFD are unlikely to be easily improved or replaced by ML. 
Many of the numerical methods underpinning CFD have been optimized for decades and provide incredible scalability and tunable accuracy.  
Imagining that ML will replace these classical numerical algorithms is unrealistic, and a more practical goal is to design new algorithms and frameworks that allow them to work together.  

Further, ML models are generally quite expensive to train, both in terms of the learning algorithm itself and the cost of generating the training data. Training data are typically generated using high-fidelity CFD, and this cost is often neglected when assessing the performance of the overall techniques, compared with traditional methods.We should also be mindful that we are comparing emerging ML techniques, many of which are less than 10 years old, with mature numerical algorithms that have been tested and developed for over half a century.  Nevertheless, it is important to include these training costs and to clarify how these models will be incorporated into existing workflows.  

A number of important questions rise when assessing an ML-based CFD solution.  How general is the model?  Does it only apply to the parameters explored in the training process, or does it generalize?  How often will the model be used, and what fidelity is required for the ultimate objective?  For example, the fidelity required for closed-loop feedback control may be considerably lower than that required to optimize an airfoil geometry with transonic separation.  
The answers to these questions inform the higher-level questions of how ML solutions will be incorporated into existing CFD workflows and how training costs should be quantified and balanced against online acceleration.  
The community is still largely in the \emph{basic research} phase of incorporating ML into scientific computing, as it remains unclear exactly how these algorithms will be used in production.  
Therefore it is important to be as precise and transparent as possible about strengths and weaknesses of these approaches.  

ML solutions, especially approaches based on deep learning, typically either require large volumes of data, or are limited to a narrow set of parameters explored in the training process.  
Improvements in transfer learning may make it possible to extend models trained at one set of parameters to another parameter set, for example generalizing an ML model to higher Reynolds numbers; however, this capability is currently lacking.  
In CFD applications, where acquiring training data requires expensive simulations, ML will likely benefit from active-learning approaches that are thoughtful in selecting training data to reduce uncertainty and improve model performance in a targeted way. There are many parallels between enhancing CFD with ML and the history of enhancing experiments with CFD, and then of enhancing CFD with ROMs. We can leverage this vast literature on uncertainty quantification, design of experiments, variable-fidelity methods and targeted idealizations to guide ML solutions.  

A promising approach to reducing training cost while improving generalizability is to incorporate known physics into the ML models.  
In fact, a good working definition of \emph{physics} is models that generalize to new unseen scenarios.  
Emerging techniques, such as physics-informed neural networks (PINNs)~\cite{Raissi2019jcp}, incorporate known physics into machine-learning techniques, for example by including a loss function to promote the governing equation being satisfied.  
While these approaches often provide improvements over training algorithms that do not incorporate any physical knowledge, adding terms to the loss function essentially acts as a \emph{suggestion} that the solution is physical.  
This is in contrast to numerical techniques that are often structurally formulated or constrained to enforce physics up to a specified numerical tolerance.  
Similarly, we should balance how much prior physical knowledge is enforced, and how rigidly, with the flexibility to capture new effects, such as model discrepancies and inadequacies.  
In the limit of training with infinite data from high-fidelity Navier--Stokes simulations, will an ML model converge to the simulation itself?  To the Navier--Stokes equations?  
These are fundamental open questions in ML, and incorporating physics is an area of intense active research.

\section*{Conclusions}\label{sec:conclusions}

Several existing, emerging, and ongoing connections are found between machine learning and computational fluid dynamics. Some challenges in CFD are particularly amenable to ML solutions, such as developing reduced-order models for large energetic coherent structures and improved RANS closure models in turbulence. In these areas, large volumes of data, often from CFD, are used to extract patterns and simplified models that may be used to accelerate future simulations, balancing accuracy and computational cost. These fields benefit from decades of experience in incorporating data-driven models and managing uncertainties through multi-fidelity methods and design of experiments.  
Thus, it is natural to incorporate machine-learning solutions into this framework.  

Other aspects of CFD are promising for augmentation through ML, although developments appear to be more challenging and improvements may be more modest.  
For example, CFD may be accelerated by learning correction terms to accurately capture high-fidelity physics on a coarser mesh, followed by a super-resolution map from the coarse mesh back to the fine mesh. 
Although initial demonstrations have been quite promising, a number of ongoing questions and challenges remain before these approaches are ready for full-scale integration with existing CFD techniques.  
Current demonstrations typically involve two-dimensional flows, and significant developments will likely be needed to extend these to more realistic three-dimensional flow configurations.  
Mature CFD techniques, such as spectral-element and finite-volume methods, are well developed and they have been optimized for several decades to have a low memory footprint and to scale to extremely large problem sizes through parallel processing.  
It may be unreasonable to expect emerging ML solutions to compete with these mature techniques without similar concerted research efforts, and time will tell how much benefit will be achieved.  
Open questions remain about the generalizability of ML models and how they will be incorporated into existing workflows, making it difficult to provide a holistic comparison of various approaches.  

Similarly, most of these techniques require extensive training databases, which are typically based on the high-fidelity CFD solutions that they aim to enhance/augment/replace.  
The cost of generating the training data is often considered a \emph{one-time, off-line} cost, with the hope of future on-line computations being accelerated. 
However, depending on the generalization capabilities of the resulting model and the number of future evaluations, the training cost may be significant and should be considered in a holistic assessment of the algorithm.  
The off-line/on-line split is especially valid for reduced-order models that will be used for control applications, where low-latency predictions are often more important than model fidelity.  
The reduced-order model literature also has extensive experience in quantifying the trade-offs between training cost, model fidelity, and on-line costs, for example to maximize the use of limited computational resources to iteratively optimize airfoil geometries for Formula One.  
This literature, which involves uncertainty quantification and iterative design of experiments, should provide a useful basis for the challenging task of incorporating ML into CFD workflows.  
We also have opportunities to improve optimization workflows, using automatic differentiation in ML frameworks to replace costly adjoint calculations.

It is important to balance optimism for the potential of ML to enhance CFD with a healthy respect for how well existing CFD works. 
We see a strong parallel between the current rise of ML and the rise of CFD in the 1980s.  
In retrospect, CFD was never destined to replace experiments, as they are complementary approaches that offer different strengths and weaknesses.  
Similarly, it should be clear that ML will not replace CFD, but will rather complement and enhance our capabilities, providing another approach to develop improved models based on available data.  
The rise of ML techniques in computational science, and in particular in CFD, will require educational efforts to incorporate these skills into the standard arsenal of graduating engineers. 
Just as engineering students now are expected to have computational proficiency, future students will be expected to have basic proficiency in ML.  

\section*{Acknowledgements}
RV acknowledges the financial support from the ERC Grant No. ``2021-CoG-101043998, DEEPCONTROL''. SLB acknowledges funding support from the Army Research Office (ARO W911NF-19-1-0045) and the National Science Foundation AI Institute in Dynamic Systems (Grant No. 2112085).

 \begin{spacing}{.7}
 \setlength{\bibsep}{1.pt}
\bibliographystyle{abbrvnat}
\bibliography{aicfd_bib}

\begin{thebibliography}{32}
\providecommand{\natexlab}[1]{#1}
\providecommand{\url}[1]{\texttt{#1}}
\expandafter\ifx\csname urlstyle\endcsname\relax
  \providecommand{\doi}[1]{doi: #1}\else
  \providecommand{\doi}{doi: \begingroup \urlstyle{rm}\Url}\fi

\bibitem[Arivazhagan et~al.(2021)Arivazhagan, Guastoni, G\"uemes, Ianiro,
  Discetti, Schlatter, Azizpour, and Vinuesa]{ari_etmm}
G.~B. Arivazhagan, L.~Guastoni, A.~G\"uemes, A.~Ianiro, S.~Discetti,
  P.~Schlatter, H.~Azizpour, and R.~Vinuesa.
\newblock {Predicting the near-wall region of turbulence through convolutional
  neural networks}.
\newblock \emph{Proc. 13th ERCOFTAC Symp. on Engineering Turbulence Modelling
  and Measurements (ETMM13), Rhodes, Greece, September 16--17. Preprint
  arXiv:2107.07340}, 2021.

\bibitem[Bae and Koumoutsakos(2022)]{bae_koumoutsakos}
H.~J. Bae and P.~Koumoutsakos.
\newblock {Scientific multi-agent reinforcement learning for wall-models of
  turbulent flows}.
\newblock \emph{Nature Communications}, 13:\penalty0 1443, 2022.

\bibitem[Benner et~al.(2015)Benner, Gugercin, and
  Willcox]{Benner2015siamreview}
P.~Benner, S.~Gugercin, and K.~Willcox.
\newblock A survey of projection-based model reduction methods for parametric
  dynamical systems.
\newblock \emph{SIAM Review}, 57\penalty0 (4):\penalty0 483--531, 2015.

\bibitem[Brunton et~al.(2016)Brunton, Proctor, and Kutz]{Brunton2016pnas}
S.~L. Brunton, J.~L. Proctor, and J.~N. Kutz.
\newblock Discovering governing equations from data by sparse identification of
  nonlinear dynamical systems.
\newblock \emph{Proceedings of the National Academy of Sciences}, 113\penalty0
  (15):\penalty0 3932--3937, 2016.

\bibitem[Brunton et~al.(2020)Brunton, Noack, and Koumoutsakos]{Brunton2020arfm}
S.~L. Brunton, B.~R. Noack, and P.~Koumoutsakos.
\newblock Machine learning for fluid mechanics.
\newblock \emph{Annual Review of Fluid Mechanics}, 52:\penalty0 477--508, 2020.

\bibitem[Choi and Moin(2012)]{choi_moin}
H.~Choi and P.~Moin.
\newblock {Grid-point requirements for large eddy simulation: Chapman’s
  estimates revisited}.
\newblock \emph{Physics of Fluids}, 24:\penalty0 011702, 2012.

\bibitem[Cranmer et~al.(2020)Cranmer, Sanchez-Gonzalez, Battaglia, Xu, Cranmer,
  Spergel, and Ho]{cranmer_et_al}
M.~Cranmer, A.~Sanchez-Gonzalez, P.~Battaglia, R.~Xu, K.~Cranmer, D.~Spergel,
  and S.~Ho.
\newblock {Discovering symbolic models from deep learning with inductive
  biases}.
\newblock \emph{Advances in Neural Information Processing Systems. Editors: H.
  Larochelle, M. Ranzato, R. Hadsell, M. F. Balcan and H. Lin}, 33:\penalty0
  17429--17442, 2020.

\bibitem[Duraisamy et~al.(2019)Duraisamy, Iaccarino, and Xiao]{duraisamy_et_al}
K.~Duraisamy, G.~Iaccarino, and H.~Xiao.
\newblock Turbulence modeling in the age of data.
\newblock \emph{Annual Review of Fluid Mechanics}, 51:\penalty0 357--377, 2019.

\bibitem[Eivazi et~al.(2021)Eivazi, Guastoni, Schlatter, Azizpour, and
  Vinuesa]{eivazi2020recurrent}
H.~Eivazi, L.~Guastoni, P.~Schlatter, H.~Azizpour, and R.~Vinuesa.
\newblock {Recurrent neural networks and Koopman-based frameworks for temporal
  predictions in a low-order model of turbulence}.
\newblock \emph{International Journal of Heat and Fluid Flow}, 90:\penalty0
  108816, 2021.

\bibitem[Eivazi et~al.(2022)Eivazi, Le~Clainche, Hoyas, and Vinuesa]{ae_modal}
H.~Eivazi, S.~Le~Clainche, S.~Hoyas, and R.~Vinuesa.
\newblock {Towards extraction of orthogonal and parsimonious non-linear modes
  from turbulent flows}.
\newblock \emph{Expert Systems with Applications}, 202:\penalty0 117038, 2022.

\bibitem[Floryan and Graham(2022)]{Floryan_Graham}
D.~Floryan and M.~Graham.
\newblock Data-driven discovery of intrinsic dynamics.
\newblock \emph{Nature Machine Intelligence}, 4:\penalty0 1113--1120, 2022.

\bibitem[Fukami et~al.(2019)Fukami, Fukagata, and Taira]{fukami2019super}
K.~Fukami, K.~Fukagata, and K.~Taira.
\newblock Super-resolution reconstruction of turbulent flows with machine
  learning.
\newblock \emph{Journal of Fluid Mechanics}, 870:\penalty0 106--120, 2019.

\bibitem[Fukami et~al.(2020)Fukami, Nakamura, and
  Fukagata]{fukami_autoencoders}
K.~Fukami, T.~Nakamura, and K.~Fukagata.
\newblock Convolutional neural network based hierarchical autoencoder for
  nonlinear mode decomposition of fluid field data.
\newblock \emph{Physics of Fluids}, 32:\penalty0 095110, 2020.

\bibitem[Guastoni et~al.(2023)Guastoni, Rabault, Schlatter, Azizpour, and
  Vinuesa]{guastoni2023deep}
L.~Guastoni, J.~Rabault, P.~Schlatter, H.~Azizpour, and R.~Vinuesa.
\newblock Deep reinforcement learning for turbulent drag reduction in channel
  flows.
\newblock \emph{The European Physical Journal E, To Appear. Preprint
  arXiv:2301.09889}, 2023.

\bibitem[Kochkov et~al.(2021)Kochkov, Smith, Alieva, Wang, Brenner, and
  Hoyer]{hoyer}
D.~Kochkov, J.~A. Smith, A.~Alieva, Q.~Wang, M.~P. Brenner, and S.~Hoyer.
\newblock {Machine learning-accelerated computational fluid dynamics}.
\newblock \emph{Proceedings of the National Academy of Sciences}, 118:\penalty0
  e2101784118, 2021.

\bibitem[Li et~al.(2020)Li, Kovachki, Azizzadenesheli, Liu, Bhattacharya,
  Stuart, and Anandkumar]{li2020neural}
Z.~Li, N.~Kovachki, K.~Azizzadenesheli, B.~Liu, K.~Bhattacharya, A.~Stuart, and
  A.~Anandkumar.
\newblock Neural operator: Graph kernel network for partial differential
  equations.
\newblock \emph{arXiv preprint arXiv:2003.03485}, 2020.

\bibitem[Lopez-Martin et~al.(2021)Lopez-Martin, Le~Clainche, and
  Carro]{sole_conv}
M.~Lopez-Martin, S.~Le~Clainche, and B.~Carro.
\newblock Model-free short-term fluid dynamics estimator with a deep
  {3D-convolutional} neural network.
\newblock \emph{Expert Systems With Applications}, 177:\penalty0 114924, 2021.

\bibitem[Milano and Koumoutsakos(2002)]{milano_koumoutsakos}
M.~Milano and P.~Koumoutsakos.
\newblock Neural network modeling for near wall turbulent flow.
\newblock \emph{Journal of Computational Physics}, 182:\penalty0 1--26, 2002.

\bibitem[Mizuno and Jim\'enez(2013)]{mizuno_jimenez}
Y.~Mizuno and J.~Jim\'enez.
\newblock {Wall turbulence without walls}.
\newblock \emph{Journal of Fluid Mechanics}, 723:\penalty0 429--455, 2013.

\bibitem[Rabault et~al.(2022)Rabault, Kuchta, Jensen, R{\'e}glade, and
  Cerardi]{rabault2019artificial}
J.~Rabault, M.~Kuchta, A.~Jensen, U.~R{\'e}glade, and N.~Cerardi.
\newblock Artificial neural networks trained through deep reinforcement
  learning discover control strategies for active flow control.
\newblock \emph{Journal of Fluid Mechanics}, 7:\penalty0 62, 2022.

\bibitem[Raissi et~al.(2019)Raissi, Perdikaris, and Karniadakis]{Raissi2019jcp}
M.~Raissi, P.~Perdikaris, and G.~Karniadakis.
\newblock Physics-informed neural networks: A deep learning framework for
  solving forward and inverse problems involving nonlinear partial differential
  equations.
\newblock \emph{Journal of Computational Physics}, 378:\penalty0 686--707,
  2019.

\bibitem[Rowley and Dawson(2017)]{Rowley2017arfm}
C.~W. Rowley and S.~T. Dawson.
\newblock Model reduction for flow analysis and control.
\newblock \emph{Annual Review of Fluid Mechanics}, 49:\penalty0 387--417, 2017.

\bibitem[Rowley et~al.(2009)Rowley, Mezic, Bagheri, Schlatter, and
  Henningson]{Rowley2009jfm}
C.~W. Rowley, I.~Mezic, S.~Bagheri, P.~Schlatter, and D.~Henningson.
\newblock Spectral analysis of nonlinear flows.
\newblock \emph{J.\ Fluid Mech.}, 645:\penalty0 115--127, 2009.

\bibitem[Schmid(2010)]{Schmid2010jfm}
P.~J. Schmid.
\newblock Dynamic mode decomposition of numerical and experimental data.
\newblock \emph{Journal of Fluid Mechanics}, 656:\penalty0 5--28, Aug. 2010.

\bibitem[Srinivasan et~al.(2019)Srinivasan, Guastoni, Azizpour, Schlatter, and
  Vinuesa]{srinivasan2019predictions}
P.~A. Srinivasan, L.~Guastoni, H.~Azizpour, P.~Schlatter, and R.~Vinuesa.
\newblock Predictions of turbulent shear flows using deep neural networks.
\newblock \emph{Physical Review Fluids}, 4:\penalty0 054603, 2019.

\bibitem[Stoll et~al.(2020)Stoll, Gibbs, Salesky, Anderson, and
  Calaf]{stoll_et_al}
R.~Stoll, J.~A. Gibbs, S.~T. Salesky, W.~Anderson, and M.~Calaf.
\newblock Large-eddy simulation of the atmospheric boundary layer.
\newblock \emph{Boundary-Layer Meteorology}, 177:\penalty0 541--581, 2020.

\bibitem[Tan et~al.(2021)Tan, Li, Lv, Yang, and Dong]{9433573}
C.~Tan, F.~Li, S.~Lv, Y.~Yang, and F.~Dong.
\newblock Gas-liquid two-phase stratified flow interface reconstruction with
  sparse batch normalization convolutional neural network.
\newblock \emph{IEEE Sensors Journal}, 21:\penalty0 17076--17084, 2021.

\bibitem[Tu et~al.(2014)Tu, Rowley, Luchtenburg, Brunton, and Kutz]{dmd_tu}
J.~H. Tu, C.~W. Rowley, D.~M. Luchtenburg, S.~L. Brunton, and J.~N. Kutz.
\newblock On dynamic mode decomposition: Theory and applications.
\newblock \emph{Journal of Computational Dynamics}, 1:\penalty0 391--421, 2014.

\bibitem[Vinuesa and Brunton(2022)]{vinuesa_brunton}
R.~Vinuesa and S.~L. Brunton.
\newblock Enhancing computational fluid dynamics with machine learning.
\newblock \emph{Nature Computational Science}, 2:\penalty0 358--366, 2022.

\bibitem[Vinuesa et~al.(2023)Vinuesa, Brunton, and McKeon]{ai4exp}
R.~Vinuesa, S.~L. Brunton, and B.~J. McKeon.
\newblock The transformative potential of machine learning for experiments in
  fluid mechanics.
\newblock \emph{Preprint arXiv:2303.15832}, 2023.

\bibitem[Wang et~al.(2019)Wang, Huang, Duan, and Xiao]{xiao_comp}
J.-X. Wang, J.~Huang, L.~Duan, and H.~Xiao.
\newblock {Prediction of Reynolds stresses in high-Mach-number turbulent
  boundary layers using physics-informed machine learning}.
\newblock \emph{Theoretical and Computational Fluid Dynamics}, 33:\penalty0
  1--19, 2019.

\bibitem[Yousif et~al.(2023)Yousif, Zhang, Yu, Vinuesa, and Lim]{heechang}
M.~Z. Yousif, M.~Zhang, L.~Yu, R.~Vinuesa, and H.~Lim.
\newblock A transformer-based synthetic-inflow generator for spatially
  developing turbulent boundary layers.
\newblock \emph{Journal of Fluid Mechanics}, 957:\penalty0 A6, 2023.

\end{thebibliography}
 \end{spacing}
\end{document}